\documentclass[
twocolumn,
superscriptaddress,
nofootinbib,
 amsmath,amssymb,
 aps,
prstper,
]{revtex4-2}

\usepackage{graphicx}
\usepackage{dcolumn}
\usepackage{bm}

\usepackage{xcolor, soul}

\begin{document}

\preprint{APS/123-QED}

\title{Attributing equity gaps to course structure in introductory physics}

\author{David J. Webb}
\affiliation{Department of Physics, University of California - Davis}

\author{Cassandra A. Paul}
\affiliation{Department of Physics \& Astronomy - Science Education Program - San Jose State University}

\date{\today}

\begin{abstract}
We add to a growing literature suggesting that demographic grade gaps should be attributed to biases embedded in the courses themselves. Changes in the structure of two different introductory physics classes were made while leaving the topics covered and the level of coverage unchanged. First, a class where conceptual issues were studied before doing any complicated calculations had zero final exam grade gap between students from underrepresented racial/ethnic groups and their peers. Next, four classes that offered students a retake exam each week between the regular bi-weekly exams during the term had zero gender gap in course grades. Our analysis indicates that demographic grade gaps can be attributed to the course structure (a Course Deficit Model) rather than to student preparation (a Student Deficit Model).
\end{abstract}

\maketitle


\section{\label{sec:Overview}Overview}
Recent research has shown that demographic gaps in introductory STEM courses correlate with demographic differences in persistence of students pursuing their STEM majors\cite{Harris2020,Hatfield2022}.  This implies that we we should be especially striving for equity  in introductory courses. However there are still some who oppose these efforts based on the perception that closing equity gaps requires lowering expectations of students. 
In his July 2022 editorial, Editor-in-Chief of \textit{Science} H. Holden Thorp recognizes this opposition to efforts aimed at allowing more underrepresented students to be successful in the sciences on the basis that these `accommodations' will diminish excellence in the field \cite{Thorp2022}. Thorp argues that ``inclusion doesn't lower standards'' by pointing out that there are many different kinds of teaching and learning methods that have been shown to allow students from different demographic backgrounds to be successful in their learning without sacrificing the quality of education. In this report, we provide additional evidence for this claim by sharing two examples of structural course changes that removed equity gaps without lowering course standards. Furthermore, we advance the discussion by providing new evidence indicating that equity gaps can't necessarily be explained by measurements of prior math and physics knowledge (i.e. a Student Deficit Model \cite{Valencia1997} may be inappropriate).  Instead, we suggest the Course Deficit Model (first discussed by Cotner and Ballen \cite{Cotner2017}) as useful when considering equity gaps.

\section{\label{sec:Framing}Research Framing}
The underrepresentation of some demographic groups in many STEM fields (see Ref's \cite{APS}, \cite{NSFStem} etc.) shows that, in these fields, those groups are denied equity in terms of access, achievement, identity, and/or power \cite{Gutierrez2008}.  In this paper we address equity in achievement, specifically achievement of underrepresented demographic groups in introductory college courses in physics.  We show evidence that demographic achievement gaps are the result of biases built into the structure of a course and may be removed by changing some features of the course. Thus, we suggest using a Course Deficit model \cite{Cotner2017} to understand these differences rather than the more commonly used Student Deficit model \cite{Valencia1997}. Using the idea that an achievement gap arises from a mismatch between course and student, the Student Deficit model looks to the detailed characteristics of the students in trying to understand the mismatch while the Course Deficit model looks to characteristics of the course to understand and close the mismatch.  In this paper we add to the growing evidence that one should look to changes in the courses themselves as a remedy for inequities in achievement between demographic groups.

The most commonly used and readily available measure of achievement is student grades and we will use such measures in this paper, using grade gaps in place of achievement gaps. In this paper we define achievement as exam or course performance as measured by grades. We suspect that there is a strong overlap between achievement and learning.  However, we are choosing to center achievement rather than learning because achievement (a student's grade) has real-life impacts regardless of associated learning. At the most basic level, grades determine if a student will need to repeat the course or not. They also provide information to advisors who advise students differently depending on those grades. A grade can also encourage or discourage a student who is deciding whether or not to stay in a certain major. 

There is a small but growing body of recent research suggesting that demographic grade gaps can be changed by changing the structure of the class in any of the following ways: 1) changing a lecture class to an active learning class \cite{Theobald2020,Ballen2017,Burke2020}, 2) changing the value of assessments in determining grades \cite{Cotner2017, Simmons2020}, and 3) changing the grade scale used to compute course grades \cite{Paul2022}. This malleability of grade gaps under changes in the structure of the class argues against the sole use of a Student Deficit model and for the inclusion of the Course Deficit model in explaining these demographic gaps.

Our analysis will also provide support for an equity model that has been called Equity of Parity \cite{Rodriguez2012}.  Following Guti\'errez \cite{Gutierrez2012}, we take demographic equity to mean that a student's achievement shouldn’t be predictable from their demographic characteristics.  Equity of Parity further includes the idea that a course should produce no demographic achievement gaps, even if there are demographic differences in measures purporting to represent the quality of a group's preparation for that course.  That is, within this equity model, \textbf{the class does not perpetuate past inequities}. 

At this point we note that finding a useful measure of student preparation for an introductory physics class that does not, itself, exhibit clear demographic biases may be difficult. First, as noted by Salehi et. al. \cite{Salehi2019}, the level of a student's previous study of physics does not explain demographic differences in college physics exam grades.  Thus, the most obvious way to prepare for a physics course is not, in their work, correlated with demographic differences in college physics grades.  Second, some other possible measures, such as SAT/ACT scores and/or FCI \cite{FCI} scores, that have been used \cite{Salehi2019,Shafer2021} to compare preparation across different demographic groups, exhibit \cite{Eberle1989,Soares2012,Dixon-Roman2013,Soares2020,Geiser2017,Traxler2018} clear demographic biases against the very demographic groups that score lower on physics exams. For instance, for many years now it has been clear that three purely demographic variables, parents' education, family income, and underrepresented group status, are significant predictors of SAT scores.  A recent analysis (in Chapter 1 of Ref. \cite{Soares2020}) by the University of California shows that these three variables alone explain an amazing 40\% of the variance in SAT scores of students applying to the university and that all three are distinctly important.  It has not been possible, to date, to prove whether these demographic biases in the scores are the result of student preparation, biases built into the SAT (or FCI), or some combination of those two factors so it becomes unclear whether these metrics measure a quantity of racism/sexism in addition to measuring a quantity of preparation.  Similarly, Madsen et al. \cite{Madsen2013} find that neither high school GPA nor prior physics experience explain the gender gap as measured by standard physics concept inventories.  For these reasons, using these metrics to control for preparation may be inappropriate in models attempting to understand the physics grade differences in different demographic groups. In other words, using ``preparation'' metrics (like the SAT) that are strongly correlated with student demographics to control for ``preparation'' may inadvertently remove visibility of any racial or gender bias that is present in the course structure itself.

This does not mean that we don't find preparation to be important but simply that measures of preparation are complex and should not necessarily be taken at face value. One example of this is that those measures that are positively correlated with achievement for students \textbf{within} a group may be differently (and even negatively) correlated with achievement in comparisons \textbf{between} groups.   Guti\'errez \cite{Gutierrez2008} suggests that the factors causing within-group differences may not be the same as those causing between-group differences. This phenomenon is observed by Shafer et al. who find that the way that we group students together impacts the the predictive power of different metrics of preparation \cite{Shafer2021}. 

Given that we are largely comparing between groups we are going to put the word ``preparation'' in quotes through the remainder of the paper because in these between-group comparisons, ``preparation'' metrics may not be measuring the same things across different demographics and therefore will not have the same predictive power as within group comparisons. By the end of this paper, we hope to provide evidence that Equity of Parity is possible even with differences in ``preparation.''

We recognize that the particular context of each class is important and that the precise changes that yield Equity of Parity for one set of students and teachers may not result in Equity of Parity with different students and teachers. Indeed, we'll see this in our data.  Nevertheless, our evidence suggests that Equity of Parity is a goal that is possible to achieve.

In this paper we describe two instances of eliminating demographic grades gaps and, importantly, also show that controlling for past ``preparation'' does not have the effects predicted by a Student Deficit model.  Thus we suggest that demographic grade gaps are determined at the course level and not the student level.  Our results provide evidence that a Student Deficit model is inappropriate, if the course can be changed, because the course organization controls essentially all of the demographic grade gap.  Along with these general conclusions we share two particular course changes that closed demographic equity gaps and so resulted in Equity of Parity being fulfilled.

\section{\label{sec:CncptFrst}Concepts-first Instruction}
First we examine some of the results of a structural change to an introductory calculus-based physics class where all of the concepts studied during a class were introduced and studied in detail in the first 60\% of the term with students working on complicated calculations only in the final 40\% of the term. We call this a ``concepts-first'' class. We compare this with the more common introductory physics classes where the various topics to be learned are studied in the same order they are arranged in the textbook. Each chapter of the text includes a discussion of the relevant conceptual material and calculations ranging from simple single-step calculations and progressing to much more complicated multi-step calculations. In a regular class these chapters are covered sequentially through the term so that there are both new concepts and new complicated calculations to learn together throughout the term. The concepts-first structured class is discussed in more detail in reference \cite{Webb2017} and in Appendix \ref{sec:CncptFrstSumm}.

Four lecture sections of this introductory physics class for physics and engineering majors were offered during the same term at a large public research university. The students from all four classes of this mechanics course took the same final exam at the same time and they were graded at the same time so we use those final exam scores to compare these two kinds of class structure. One of the chapter-by-chapter classes (Section II)  and the concepts-first class (Section I) were taught by the same instructor using the same lecture slides, student activities, and homework problems but with a different timing of the various parts of the course to make one class chapter-by-chapter and one class concepts-first.  Both of these classes can be considered active-learning classes in that much of the lecture time was spent in student-student discussions of conceptual ideas.  The other two chapter-by-chapter classes (Sections III and IV) were taught by veteran instructors who had taught the course many times and whose courses were more traditional. The students registered for the various classes without any foreknowledge of how the classes would be organized. 

We examine the grade gaps of the demographic groups that the American Physical Society identifies \cite{APS} as underrepresented in physics: i) racial or ethnic background (we use the acronym URM to identify students with either African, Hispanic, Indigenous American, and/or Pacific Islander ethnicity) and ii) gender (APS uses binary gender and identifies female students as underrepresented).  We use the university supplied data on the students’ self-identified racial/ethnic and binary gender categories. At the time these data were collected, the university only recognized two genders that matched those assigned at birth. We regrettably have no means to collect more accurate gender information.

The final exam was scored out of 160 points which is an awkward number so we normalize the final exam grades so that average over all 633 students is zero with a standard deviation equal to one. We chose to do this to make the units more understandable, but importantly, if we instead use the raw scores, our results are the same.  When we compare the average final exam grade of URM students with the average grade of their peers in the same class, the units will be standard deviations. The grading of each exam problem was done by the same people for all four classes so as to eliminate any possible differences in grading.

We separately consider the results using a Course Deficit model and a Student Deficit model.  For the latter model, as discussed in Ref. \cite{Webb2017}, we use two measures of student ``preparation'' as they entered the class, a survey of physics concepts \cite{FCI} and the students' normalized introductory calculus grades.  The student demographics of our final database including all of these data are shown below in Table \ref{tab:tab1}.

\begin{table}[htbp]
\caption{Demographics of the four lecture sections of this introductory course in Newtonian mechanics included in the dataset.}
\label{tab:tab1}
\begin{tabular}{c c c c }
\textbf{Section} & \textbf{N} & \textbf{\%URM} &
\textbf{\%Female} \\ 
\hline 
I & 152 & 13 & 25 \\
II & 160 & 13 & 24 \\
III & 163 & 22 & 31 \\
IV & 158 & 10 & 22 \\
\end{tabular}
\end{table}

\begin{figure}
\includegraphics[scale = 0.5, trim=5.5cm 2.4cm 5.2cm 3.6cm, clip=true,width=8.6cm]{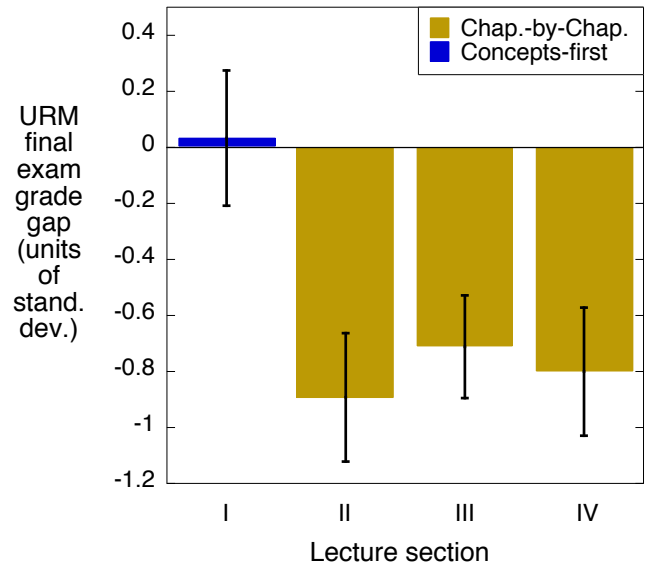}
\caption{The URM grade gap on the final exam is the URM average final exam score minus the average final exam score of their peers in the same class.  The final exam distribution is normalized to standard deviation = 1 and the results for the four different classes taking the same final exam are shown.  Classes II, III, and IV were taught chapter by chapter in the usual way and class I was taught concepts first.  Class I and II were taught by the same instructor using exactly the same materials (lecture slides, student activities, and homework) but just arranged differently in time in the two classes.  The error
bars are standard errors.}
\label{Fig1}
\end{figure}

For our first analysis, we do not attempt to control for students' prior ``preparation'' because we want to see the impact that the concepts-first course has on closing grade gaps in general.  The differences in the average final exam grades of URM and non-URM students are 0.03 $\pm$ 0.24, -0.89 $\pm$ 0.23, -0.71 $\pm$ 0.18, -0.81 $\pm$ 0.23 for lectures I, II, III, and IV, respectively and these are plotted in Figure \ref{Fig1} for each of the four classes.  A negative gap means the URM average was lower than non-URM. There is a distinct negative grade gap for each of the three classes (II through IV) taught chapter-by-chapter with the URM students having lower average grades. These grade gaps are roughly equal to each other.  In addition, they are also comparable to grade gaps published by several other US universities \cite{Salehi2019,Shafer2021} in that they are all negative and a fraction of a standard deviation, even though the actual exams given in these other schools are likely very different. On the other hand, in the concepts-first class (class number I) URM students had slightly higher final exam grades than their peers though the result is consistent with zero gap.

To analyze the differences seen in Fig. \ref{Fig1}  we first group together the three traditional classes.  From here on, each time we group more than one class in a single analysis we will do that using Hierarchical Linear Modeling (HLM) with STATA software.  We use HLM to account for the fact that there are class-to-class differences in the exact material that students worked on and studied during the quarter and class differences such as these are expected to lead to class-level correlations on the final exam.  For instance, students from section IV had seen two of the final exam problems (and their solutions) during the quarter as well as part of another question, students from section III had seen one final exam problem (and its solution) and also saw the same exam layout on two midterms as they had for the final, and there are likely other class differences that we don't know about but that can affect the exam results.  HLM models each class by itself before assembling those results into the final coefficients so it should account for differences in the lecture sections because URM and non-URM students in the same section saw the same course materials.  Nevertheless, as we show in Appendix \ref{sec:OLS}, essentially none of our results would change appreciably if we had instead simply used the more common ordinary least-square (OLS) fitting. For a discussion of HLM see Ref. \cite{VanDusen2019}.

\begin{figure}
\includegraphics[scale = 0.5, trim=5.0cm 2.4cm 5.2cm 3.6cm, clip=true,width=8.6cm]{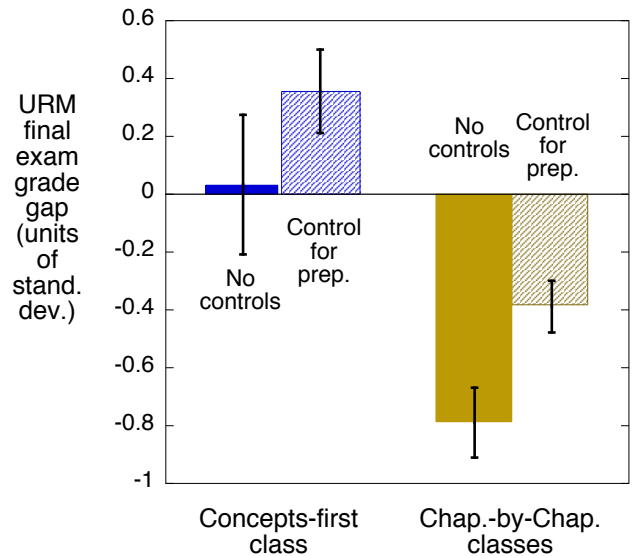}
\caption{Comparing the concepts-first class with the three chapter-by-chapter classes grouped together.  Again, the URM grade gap on the final exam is positive if URM students outperformed their peers.  For each class organization the bare uncontrolled URM gap is shown as well as the URM gap after controlling for incoming math and physics understandings of the students. The error bars are standard errors.}
\label{Fig1a}
\end{figure}

In modeling the normalized exam grade ($NFnlExam$) we define two categorical variables.  At the student level, $URM = 1$ if the student identified their ethnicity as placing them in the URM category and $URM = 0$ if they didn't.  At the class level, $CncptFrst = 1$ if a student is in the concepts-first class (Section I) and $CncptFrst = 0$ if they were enrolled in one of the other sections.  First, we fit the following model separately for the two types of class:
\begin{equation}
NFnlExam = b_0 + b_{URM}URM \label{eqn:HLMModel0}
\end{equation}
This analysis yields a URM gap for the three chapter-by-chapter classes of $b_{URM} = -0.79 \pm 0.12$ and, of course, there is only one concepts-first class so that gap is the same one we found above.  These numbers for the gaps uncontrolled for ``preparation'' are plotted in Figure \ref{Fig1a}.

Next we use HLM to give us a numerical comparison of the concepts-first class to the chapter-by-chapter classes.  The model we fit includes both $URM$ and $CncptFrst$ and the interaction between them:
\begin{multline}
NFnlExam = b_0 + \\ b_{CncptFrst}CncptFrst + b_{URM}URM + \\ b_{URM*CncptFrst}(URM*CncptFrst)
\label{eqn:HLMModel1}
\end{multline}

The results of our HLM fit to equation \ref{eqn:HLMModel1} are shown in Table \ref{tab:tab2}.  From $b_{CncptFrst}$ we see that the non-URM students from the concepts-first class had final exam grades that were statistically indistinguishable from students from the regular classes (despite the fact that some regular lecture sections had seen some exam problems during the term).  Second, $b_{URM*CncptFrst}$ is significantly different from zero so the URM students from the concepts-first class did much better on the final exam than their URM peers in the regular classes.  Finally, $b_{URM}$ is the demographic grade gap found in the regular classes. So, the grade gap in the concepts-first class is about 3.16 standard errors above the background gap seen in the chapter-by-chapter classes.  This suggests that there is about one chance in 500 that this difference is simply a random fluctuation (i.e. $P=0.002$). Because teaching concepts first removes the equity gap that exists in the chapter-by-chapter class, this is evidence in favor of using a Course Deficit model in understanding the URM gaps in this set of classes. 

\begin{table}[htbp]
\caption{The coefficients from an HLM fit to equation \ref{eqn:HLMModel1} are shown along with their standard errors, z-statistics, and P-values.  Included are $N=633$ students in 4 classes.  The interaction term suggests that the URM gap is significantly different (reduced) for the concepts-first class.}
\label{tab:tab2}
\begin{tabular}{c c c c c}
\textbf{Coeff.} & \textbf{Value} &\textbf{Error} & \textbf{z-statistic}
& \textbf{P-value}\\ 
\hline
$b_{CncptFrst}$ & -0.19 & 0.20 & -0.94 & 0.345 \\
$b_{URM}$ & -0.79 & 0.12 & -6.41 & $<10^{-3}$ \\
$b_{URM*CncptFrst}$ & 0.82 & 0.26 & 3.16 & 0.002 \\
$b_{0}$ & 0.15 & 0.10 & 1.45 & 0.147 \\
\end{tabular}
\end{table}

If the URM gaps were explainable in terms of student preparation (a Student Deficit model) then controlling for that ``preparation'' should shrink each gap, and the difference between the two class types, to zero. We use the students' normalized calculus grades, $Calc$, along with the Force Concept Inventory survey, $PreFCI$,  to control for their incoming math and physics ``preparation'' in an HLM analysis of the URM final grade gaps.  In other words, we fit the normalized final exam scores with the following model:
\begin{multline}
NFnlExam = b_0 +\\ b_{Calc}Calc + b_{PreFCI}PreFCI +\\ b_{URM}URM
\label{eqn:HLMModel1a}
\end{multline}
The results of using this model on each of the two types of class are that the URM grade gaps, $b_{URM}$, are $-0.388 \pm 0.089$ for the group of three chapter-by-chapter classes and $0.36 \pm 0.14$ for the concepts-first class. These numbers are also plotted in Figure \ref{Fig1a}.  Neither gap is consistent with zero after using this Student Deficit model and the estimated gap for the concepts-first class has increased instead of decreasing.

We can put all four classes into the same model using:
\begin{multline}
NFnlExam = b_0 + \\b_{Calc}Calc + b_{PreFCI}PreFCI +\\ b_{CncptFrst}CncptFrst + b_{URM}URM +\\ b_{URM*CncptFrst}(URM*CncptFrst)
\label{eqn:HLMModel2}
\end{multline}

The results of our HLM fit to Equation \ref{eqn:HLMModel2} are shown in Table \ref{tab:tab3}.  Again, $b_{CncptFrst}$ is small and statistically insignificant so, again, we see that the non-URM students performed essentially equally in the two kinds of class organizations.  However, $b_{URM*CncptFrst}$ again shows us that the URM students in the concepts-first class had final exam scores over 4 standard errors above the background (chapter-by-chapter) classes. This analysis shows that the Student Deficit model does not appear to help us at all in explaining the URM grade gap differences seen in the different class organizations. Controlling for ``preparation'' in the Chapter-by-Chapter class does explain some of the gap, but the same ``preparation'' metric does not explain the gap in the concepts-first class. The metrics of student preparation are not always correlated with final exam grades in the same way.

\begin{table}[htbp]
\caption{The coefficients from an HLM fit to equation \ref{eqn:HLMModel2} are shown along with their standard errors, z-statistics, and P-values.  Included are $N=633$ students in 4 classes.}
\label{tab:tab3}
\begin{tabular}{c c c c c}
\textbf{Coeff.} & \textbf{Value} &\textbf{Error} & \textbf{z-statistic}
& \textbf{P-value}\\ 
\hline
$b_{Calc}$ & 0.608 & 0.041 & 14.68 & $<10^{-3}$ \\
$b_{PreFCI}$ & 0.0692 & 0.004 & 16.77 & $<10^{-3}$ \\
$b_{ConcptFrst}$ & -0.10 & 0.17 & -0.55 & 0.584 \\
$b_{URM}$ & -0.378 & 0.087 & -4.35 & $<10^{-3}$ \\
$b_{URM*ConcptFrst}$ & 0.73 & 0.18 & 4.07 & $<10^{-3}$ \\
$b_{0}$ & -1.19 & 0.11 & -11.02 & $<10^{-3}$ \\
\end{tabular}
\end{table}

Finally, we note that HLM analysis also shows that the concepts-first class and the traditional classes had about the same size gender gap (see Appendix \ref{sec:OtherGaps}) and the same grade gap for Asian students (see Appendix \ref{sec:Asian}).  Again, all four courses covered the same material at the same level using the same textbook and taking the same final exam. Furthermore, this is not obviously an instructor effect because the instructor who taught the concepts-first class also taught one of the chapter-by-chapter classes.

\section{\label{sec:Retakes}Assessment Retakes Course}
Second, we examine the results of a change in the assessment structure of an introductory series of physics courses for biological science students. Calculus is required as a prerequisite for this course, but the course is mainly algebra-based. All of the courses considered here are active-learning classes (these classes were offered at the same public research university as the concepts-first class) and are discussed in some detail in Ref. \cite{Potter2014Sixteen}. These classes generally have one 80 minute lecture and two 140 minute discussion/labs per week. The students in these classes usually take either one 20 minute exam on new material every lecture or one exam on new material every two lectures, and a final exam at the end. The assessment structure of a class was changed in four classes over three terms. In these classes students had one exam on new material every other lecture and in the intervening lectures an optional ``retake'' exam was administered that covered the same material and could supplant the original grade if the retake score was higher \cite{Webb2021}.  On retake day the students could stay to take a retake exam in the final 25 minutes of class or leave class early if they felt they had done well enough on the original exam.  In addition, the students could look at and/or work on the retake and then decide to leave without turning it in.  No retake was possible for the final exam in either type of class. In both the non-retake classes and the retake classes the course grade was almost entirely determined by exam scores (the four short exams + final exam) \cite{Webb2020} because these classes do not grade homework. This allows us to compare course grades as a proxy for exam scores. We did not measure student initial understandings of physics or math but we can use a student’s incoming GPA as a control variable to serve as a proxy for their general academic ability. 

The course grade distributions in these courses have an average standard deviation of about one grade point so a course grade gap size in this course will have a meaning similar to the normalized final exam gap sizes discussed above for a different introductory physics series. We compare the 4 retake classes with a baseline computed from all 52 of the classes offered within this course series over the 2.5 years immediately previous to the first retake course. The demographics of our final database are shown below in Table \ref{tab:tab4}.

\begin{table}[htbp]
\caption{Demographics of the 56 lecture sections of this introductory physics series included in the dataset.}
\label{tab:tab4}
\begin{tabular}{c c c c}
\textbf{Type} & \textbf{N} & \textbf{\%URM} &\textbf{\%Fem.} \\ 
\hline 
Retake & 610 & 23 & 66 \\ NonRet. & 12,884 & 18 & 63 \\
\end{tabular}
\end{table}

\begin{figure}
\includegraphics[scale = 0.5, trim=4.5cm 2.0cm 5.5cm 3.5cm, clip=true,width=8.6cm]{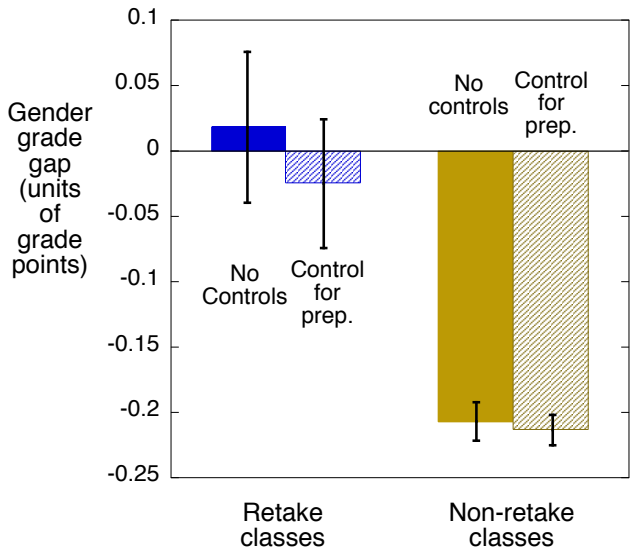}
\caption{The gender gap for the course grade is the average course grade for female students minus the average course grade for male students in the same class.  The course grade distribution already has a standard deviation of about one so it was not normalized.  These classes all had course grades largely determined by exam grades and the classes that allowed retake exams had little or no gender grade gap.  The error bars are standard errors.}
\label{Fig2}
\end{figure}

We use HLM to find the average, over classes, of the difference between course grades of female students and course grades of male students in the same class. In practice this means that we fit
\begin{equation}
CourseGrade = b_0 + b_{Female}Female \label{eqn:HLMModel3}
\end{equation}
where $Female = 1$ if the student identified as female in our database and 0 if they identified as male and $CourseGrade$ is the grade the student received in the course. Figure \ref{Fig2} shows the differences of these two average grades ($b_{Female}$) for 2.5 years of these classes, immediately preceding the first retake class, to identify the non-retake gender grade gap, and the gender grade gap in the four courses that allowed retakes. Female students had lower average course grades than male students in the non-retake courses.  Again, these grade gaps are comparable to grade gaps published by other US universities \cite{Salehi2019} in that they are the negative of a fraction of a standard deviation. 

On the other hand, in the retake classes female students had slightly higher course grades than male students. To quantify the comparison between the 52 non-retake classes and the 4 retake classes we define the categorical variable $Retake$ =1 for the classes that offered retake exams and = 0 for the classes that did not offer retake exams.  We use HLM to fit
\begin{multline}
CourseGrade = b_0 + b_{Retake}Retake \\+ b_{Female}Female \\+ b_{Female*Retake}(Female*Retake)
\label{eqn:HLMModel4}
\end{multline}
The results of our HLM fit to equation \ref{eqn:HLMModel4} are shown in Table \ref{tab:tab5}.  From $b_{Female}$ we find the gender gap we already knew about from the regular courses.  From $b_{Retake}$ we find that students identifying as male had about 1/3 of a grade point higher grades under the retake grading regime.  Finally, from $b_{Female*Retake}$ we find that female students in the retake classes had an additional 1/4 of a grade point so that the gender gap is essentially gone.

\begin{table}[htbp]
\caption{The coefficients from an HLM fit to equation \ref{eqn:HLMModel4} are shown along with their standard errors, z-statistics, and P-values.  Included are $N=12,884$ students in 52 non-retake classes and $N = 610$ students in 4 retake classes.}
\label{tab:tab5}
\begin{tabular}{c c c c c}
\textbf{Coeff.} & \textbf{Value} &\textbf{Error} & \textbf{z-statistic}
& \textbf{P-value}\\ 
\hline
$b_{Retake}$ & 0.31 & 0.14 & 2.18 & 0.029 \\
$b_{Female}$ & -0.210 & 0.015 & -13.75 & $<10^{-3}$ \\
$b_{Female*Retake}$ & 0.229 & 0.073 & 3.15 & 0.002 \\
$b_{0}$ & 3.076 & 0.036 & 84.67 & $<10^{-3}$ \\
\end{tabular}
\end{table}
The average grade gap in the retake classes is about 3.15 standard errors above the background gap seen in the regular classes (i.e. $P = 0.002$).  This suggests that there is only about one chance in five hundred that this difference is just a random fluctuation and is evidence that a Course Deficit model, again, is appropriate to use in understanding the demographic gender gap in the normal courses.

Now we control for the students' demonstrated academic abilities using their incoming GPA in case the classes giving retake exams had female students who were much better students than their male peers.  First, we fit Eq. \ref{eqn:HLMModel5a} for the two groups of classes, retake and non-retake, separately to find gender gaps, $b_{Female}$, of $-0.025 \pm 0.049$ for the retake classes and $-0.214 \pm 0.012$ for the non-retake classes after controlling for incoming GPA.  These GPA-controlled gaps are plotted in Fig. \ref{Fig2} next to the uncontrolled gender gaps and we see that controlling for GPA doesn't change the gaps much.
\begin{multline}
CourseGrade = b_0 + b_{GPA}GPA + b_{Female}Female
\label{eqn:HLMModel5a}
\end{multline}

Now we put the two types of class into the same model to quantify the difference after controlling for GPA as follows:
\begin{multline}
CourseGrade = b_0 + b_{GPA}GPA \\+ b_{Retake}Retake + b_{Female}Female \\+ b_{Female*Retake}(Female*Retake)
\label{eqn:HLMModel5}
\end{multline}

The results of our HLM fit to equation \ref{eqn:HLMModel5} are shown in Table \ref{tab:tab6}.  The difference, $b_{Female*Retake}$, is still significantly different from 0 and continues to suggest that the gender gap is an artifact of the structure of the course.
\begin{table}[htbp]
\caption{The coefficients from an HLM fit to equation \ref{eqn:HLMModel5} are shown along with their standard errors, z-statistics, and P-values.  Included are $N=12,884$ students in 52 non-retake classes and $N = 610$ students in 4 retake classes.}
\label{tab:tab6}
\begin{tabular}{c c c c c}
\textbf{Coeff.} & \textbf{Value} &\textbf{Error} & \textbf{z-statistic}
& \textbf{P-value}\\ 
\hline
$b_{GPA}$ & 1.108 & 0.012 & 89.12 & $<10^{-3}$ \\
$b_{Retake}$ & 0.38 & 0.15 & 2.53 & 0.011 \\
$b_{Female}$ & -0.214 & 0.012 & -17.68 & $<10^{-3}$ \\
$b_{Female*Retake}$ & 0.174 & 0.058 & 3.02 & 0.003 \\
$b_{0}$ & -0.320 & 0.055 & -5.81 & $<10^{-3}$ \\
\end{tabular}
\end{table}
So the difference in the gaps between the two classes is not significantly decreased if one uses incoming GPA to control for the students’ academic ability with the error estimate substantially the same, the retake classes are about 3.02 standard errors above the background set by the regular classes ($P = 0.003$) so it is not obvious that a Student Deficit model is of any use in understanding these differences. In other words, even though incoming GPA is a significant predictor for individual success in the course, controlling for this at the individual level does not significantly change the gender gap in in either set of courses.

Finally, as we show in Appendix \ref{sec:OtherGaps}, the retake classes and the non-retake classes had about the same size URM demographic gap, with URM students receiving lower average grades than non-URM students under each assessment regime.  So this particular intervention does not appear to close the URM demographic gap.  Also note, we found (see $b_{Retake}$ in Table \ref{tab:tab5}) that male students had higher average grades in the retake classes than they had in the regular classes. In Appendix \ref{sec:TwoInstr} we show that \textbf{each} of the two instructors teaching the retake classes had significant non-zero gender gaps in these courses when they taught them without retakes and that \textbf{each} of the instructors had gender gaps consistent with zero when they taught their retake courses.  Again, we note that the retake classes and the non-retake classes covered the same material at the same level and with approximately the same course materials.

\section{\label{sec:Discussion}Discussion}

Several recent PER papers make measurements similar to those above and may be viewed through a Student Deficit vs Course Deficit lens (even though the authors do not originally use those terms). A recent study by Salehi et al. \cite{Salehi2019} analyzes student performance with respect to preparation and concludes that \emph{``when controlling for incoming preparation, there remain no}
[significant] \emph{demographic performance gaps.''} Salehi et al. argue that average deficits in the preparation of some demographic groups of students explain a substantial portion of the exam achievement gap that these groups experience under the particular unspecified teaching/assessment regimes of three different universities. They also suggest \emph{``It is possible
that there is some unmeasured factor (e.g., test anxiety) that
causes both lower scores on our measures of incoming
preparation and lower final exam performance.''}  This perspective can be viewed as a Student Deficit model - that individual student preparation (or some other student-level variable) is responsible for the equity gaps. We offer an alternative explanation using a Course Deficit model; since Salehi et al. quantify ``preparation'' using measurements that, as we have noted earlier, are themselves suspected of including biases against the relevant demographic groups, the courses and/or course exams are potentially subject to those same biases. Therefore controlling for one bias removes the other. Alternatively, two other recent papers, by Shafer et al. \cite{Shafer2021} and Stewart et al. \cite{Stewart2021}, use similar metrics of student preparation in the same kinds of calculations used by Salehi et al. but conclude that student preparation does not explain the various achievement gaps they discuss. Shafer et al. find that preparation metrics do not predict student success equally across demographic groups. It's clear they are using a Course Deficit model as they conclude that \emph{``There may be
something about the physics course, the engineering
program, or student culture that prevents Asian
American and African American students, and to a lesser
extent, Hispanic students, from realizing their full potential''} \cite{Shafer2021}.

These papers may be suggesting that, because a Student Deficit model does not explain achievement gaps, a Course Deficit model is needed. Again, our view is that ``preparation'' is very difficult to measure when comparing different demographic groups. Burkholder et al. \cite{Burkholder2022} reports that providing extra help to students who entered with purported ``preparation deficits'' did not close the achievement gaps. In our view adopting a Student Deficit model tends naturally to lead one to the idea of giving some students extra preparation to decrease the ``preparation deficit''.  After pointing out that they tried this and it failed to decrease the gaps, Burkholder et al. \cite{Burkholder2022} seem to adopt a Course Deficit model as most of the paper is concerned with changes they made to their courses and the results of these changes.  Unfortunately, their measure of equity is not clearly related to the demographic achievement gaps that we have discussed above as their definition of equity addresses ``preparation'' gaps without examining demographic gaps that might exist. Nevertheless, their work agrees with our main conclusion - that there needs to be increasing focus on introductory courses, themselves, as the causes of demographic gaps. One final application of a Course Deficit model involves looking at the characteristics of courses that are correlated with larger equity gaps.   Canning et al. \cite{canning_stem_2019} find, and Park et al. \cite{Park2021} replicate, that STEM courses taught by faculty with fixed mindsets have  equity gaps that are twice as large on average as those taught by faculty with growth mindsets. Use of a Course Deficit model in this case would imply that faculty with fixed mindsets are more likely to employ course structures that, for whatever reason, increase equity gaps. 

Our work suggests that the Course Deficit model may be all that is needed in explaining the grade gaps and that a Student Deficit model may be inappropriate for these issues. We might reconcile these various ideas by suggesting that if i) introductory physics courses were unchangeable for some reason or ii) changing around the structure of introductory physics courses always led to the same rough demographic gaps then a Student Deficit model would be appropriate. However, the structure of physics courses may be changed and the data discussed in this paper show that demographic gaps are not only changeable but may sometimes even change sign. 

In our introduction we noted that there are good reasons to worry that measures such as SAT/ACT math scores and FCI scores are biased against some demographic groups \cite{Madsen2013, Eberle1989,Soares2012,Dixon-Roman2013,Soares2020,Geiser2017,Traxler2018} so that using these measures to compare different demographic groups may be inappropriate.  Now we argue that the data in our paper together with the data in other published research are consistent with that conclusion, so that these measures of preparation should probably be used only for within demographic group comparisons. In many of the cases discussed in the literature \cite{Salehi2019,Shafer2021,Stewart2021} an underrepresented group received a lower average grade than their peers and this negative achievement gap is reduced (i.e. the gap changes in the positive direction) after controlling for math and physics ``preparation'' scores.  This kind of change in a negative gap would occur whether the math and physics ``preparation'' scores measured bias against the underrepresented group or whether they measured poorer preparation of that group.  So, most of these data don't help us decide whether the measures are controlling for bias or control for preparation. However, with the data from this paper there are now at least two results that differ from this norm of a negative gap becoming less negative after controlling for math and physics ``preparation''.  One uncommon result was shown above in Fig. \ref{Fig1a} where a positive URM gap became even more positive after controlling for math and physics ``preparation''. A surprising result like this is inconsistent with these measures acting as controls for preparation but is consistent with them acting as controls for bias against URM students. A second uncommon result was seen in Ref. \cite{Shafer2021} which showed that Asian-American students had a negative exam grade gap but that this negative gap became larger after controlling for math and physics ``preparation''.  This result is also inconsistent with the idea that these control variables measure preparation. These results suggest that perhaps these metrics should not be considered as proxies for preparation.






As a caution, we also find that any single change in course structure may differentially benefit one underrepresented demographic group and not benefit other demographic groups. Added to this caution is our personal teaching experience that these issues are at least somewhat dependent on the particular teacher, the particular group of students, and, for each teacher and student, can change from term to term.  Unfortunately, research done in support of physics education has primarily been done at institutions that have majority white students with above average SAT scores \cite{Kanim2020} so it is hard to know what might be applicable in a particular class.  We also note that the concepts-first study lasted only one academic term and covered only Newtonian mechanics.  In general we suggest that a teacher be conscious of their student populations and any existing equity gaps when making choices about their course design. The knowledge-base that teachers can draw on in structuring their courses is, unfortunately, not very complete. There are likely many many possible ways to restructure traditional courses beyond those discussed in the literature or in this paper so more research is necessary on the differential demographic impacts of different course structures. Finally, there are also issues with the definitions of the demographic groups themselves; for instance, assuming gender is binary or aggregating several ethnicities into a single group \cite{Shafer2021,Paul2022} or ignoring the intersectional nature inherent in the definitions of these groups. These are limitations in our own work, and they should be more broadly investigated.

When considering these results, it's also important to remember that eliminating achievement grade gaps does not necessarily eliminate all equity gaps from the course. As we note in our introduction, other equity gaps still might exist - particularly those that are related to access, identity and power. Also, while we have relied on our data to advocate against using a Student Deficit model, we also note another argument against using a Student Deficit model, using such a model can potentially perpetuate the same racist and sexist perspectives \cite{Valencia1997} responsible for the gaps in the first place.

\section{\label{sec:Conclusions}Conclusions}

We summarize three main conclusions from our analysis: 1) We find two examples of course changes that successfully eliminated some demographic grade gaps when compared with a control group: a) teaching concepts first resulted an equity gap consistent with zero for underrepresented minority students, and b) allowing retake exams resulted in an equity gap consistent with zero for female students; 2) Equity of parity was achieved for these demographic groups without controlling for any incoming inequities and (importantly) without changing the academic standards of the course. 3) When we did control for incoming inequities (often discussed as ``preparation'' metrics in the literature), those metrics did not reduce the grade gaps in predictable ways. Because controlling for individual student ``preparation'' did not reduce the equity gaps, we argue that  grade gaps are the result of the course (a course deficit model) and not the individual students (a student deficit model).

This paper adds to a growing literature  that changes in the structure of a course, without changing the course content or the level of content expertise expected by the STEM community, may affect different demographic groups differently \cite{Theobald2020,Ballen2017,Burke2020,Cotner2017, Simmons2020, Paul2022}.  These changes were, initially, done in an attempt to benefit \textbf{all} of the students in the class and in this paper, together with Ref. \cite{Webb2017}, they have been shown to do that. But we see that they also benefit some demographic groups more than others. Therefore because demographic grade gaps seem to be quite changeable under changes in course structure without applying interventions to address any existing student-level ``preparation'' gaps present at the beginning of the course, it seems wisest to use such measures simply to judge one course structure against another rather than one group of students against another. Furthermore, because we find that controlling for metrics used to describe ``preparation'' can either decrease or increase the demographic grade gaps (as we see in Figure \ref{Fig1a}) depending on the course context, these metrics should perhaps not be used so readily to explain grade gaps.  In other words, our data support using a Course Deficit model of demographic grade gaps rather than a Student Deficit model. Taken all together, these ideas also support the idea that Equity of Parity is an appropriate goal for all introductory physics classes and, perhaps, for all STEM classes. An Equity of Parity model also supports a goal that many teachers may have, the goal of \textbf{not perpetuating past inequities.}

We conclude that there are likely systemic biases, in introductory physics classes, that act against some underrepresented demographic groups. These biases are easily seen by comparing outcomes between different systems of teaching and assessment and these biases can likely be removed with appropriate structural changes at the level of a course that -importantly- do not impact the educational standards of the course.

\begin{acknowledgments}
We appreciate Rod Cole, Richard Scalettar, and John Terning for helping us organize these studies. 
We also thank the San Jose State University PER group for reviewing and providing feedback on an early draft of this paper.  Finally, we are indebted to Wendell Potter (deceased Jan. 2017) who provided mentorship throughout both studies.  
\end{acknowledgments}

\appendix

\section{\label{sec:CncptFrstSumm}Short intro to Concepts-first class}
Given that we would like students to learn the conceptual foundations of physics and use these concepts while learning computational skills, a decision was made to try teaching a class where the first part of the term is largely conceptual so that students were as well-grounded as possible in the conceptual ideas before they use those ideas in honing their skills at computation.  To that end the first 60\% of the term in our concepts-first class asked the students to understand the main ideas by using these ideas in analyzing both simple and complicated physical situations using the appropriate words, graphs, diagrams, pictures, and equations to identify and describe the relations between the relevant physical variables.  During this first part of the term discussions of the simplest physical situations may end with a simple (single equation) calculation but discussions of the more complicated physical situations (those whose solution requires more than one scalar equation or those that cannot even be solved by students in this level of class) are stopped after the situation has been examined using words, graphs, and/or diagrams and before writing down the more complicated equations.  After the students have had many chances to understand the applications of the term’s physical ideas, the final 40\% of the term is spent practicing applying these same ideas in more complex computations.  Of course doing these complex calculations necessarily involves reviewing the ideas from the entire term.  In the first 60\% of the term a typical lecture consisted of a description of the ideas using words, examples, and equations followed by “clicker questions” asking the students to make their own sense of these ideas using words, pictures, and graphs with the relevant physical variables.  During this first 60\% of the term the discussion time was similarly spent on small-group work in understanding the conceptual issues involved in a variety of physical situations, the homework problems were like the discussion problems, and the exams were, in the same sense, conceptual.  The final 40\% of the term included lecture, discussion, and homework on the more complicated computational problems.  The course materials are available online with Ref. \cite{Webb2017}.

\section{\label{sec:Incoming}Incoming Variables from Concepts-First}
The first concepts-first paper\cite{Webb2017} had a demographic breakdown of the incoming academic characteristics of the students but Sections III and IV were grouped together.  For completeness we will give the incoming characteristics of the students included in this paper in Table \ref{tab:tab1a}.  Our paper shows that these small demographic differences in incoming characteristics do not seem to be connected with demographic differences in outcomes.

\begin{table}[htbp]
\caption{Incoming averages and standard deviations (in parentheses) for measured math and physics understandings for each of the four lecture sections of this introductory course in Newtonian mechanics. The averages are over i) URM students and ii) Non-URM students separately.  Our paper suggests that these differences are irrelevant to outcomes under the appropriate course structure.}
\label{tab:tab1a}
\begin{tabular}{c c c c}
\textbf{Class} & \textbf{Group} & \textbf{Calculus} & \textbf{pre-FCI} \\
\hline
Lecture & URM & -0.20 (0.74) & 15 (6) \\
I & Non-URM & 0.22 (0.67) & 16 (7)  \\
\hline
Lecture & URM & -0.11 (0.49) & 14 (6) \\
II & Non-URM & 0.13 (0.68) & 17 (6) \\
\hline
Lecture & URM & 0.16 (0.72) & 14 (6) \\
III & Non-URM & 0.50 (0.64) & 17 (7)  \\
\hline
Lecture & URM & -0.27 (0.49) & 13 (7) \\
IV & Non-URM & 0.45 (0.65) & 15 (6)  \\
\hline
\end{tabular}
\end{table}

\section{\label{sec:OLS}Ordinary Least Square Fits}
In this appendix we show how the results can differ if we aggregate the data and fit a model using an ordinary least squares (OLS) procedure.  We can use the model from Equation \ref{eqn:HLMModel1} to show how OLS fitting to the model mostly just reproduces the HLM results found in Table \ref{tab:tab2} but, in addition, incorrectly treats the lecture-level variable, $ConcptFrst$.  Using OLS to fit Equation \ref{eqn:HLMModel1} yields the coefficients shown in Table \ref{tab:tab7}.
\begin{table}[htbp]
\caption{The coefficients from an OLS fit to equation \ref{eqn:HLMModel1} are shown along with their standard errors, t-statistics, and P-values.  Included are $N=633$ students in 4 classes.}
\label{tab:tab7}
\begin{tabular}{c c c c c}
\textbf{Coeff.} & \textbf{Value} &\textbf{Error} & \textbf{t-statistic}
& \textbf{P-value}\\ 
\hline
$b_{ConcptFrst}$ & -0.19 & 0.10 & -1.96 & 0.050 \\
$b_{URM}$ & -0.79 & 0.12 & -6.38 & $<10^{-3}$ \\
$b_{URM*ConcptFrst}$ & 0.83 & 0.26 & 3.13 & 0.002 \\
$b_{0}$ & 0.15 & 0.05 & 3.04 & 0.002 \\
\end{tabular}
\end{table}
Comparing Table \ref{tab:tab2} with Table \ref{tab:tab7}, we see that the estimated error values for $b_{ConcptFrst}$ are smaller than when using HLM.  OLS treats this variable as independently varying over students.  In reality this variable only varies over classes as all students in any particular lecture class have exactly the same value of $ConcptFirst$.  Treating this as a student level variable will certainly lead to the error estimate being lower than it should be and we find it reduces the error estimate by about 50\%.  For our purposes, the important variables and their error estimates are basically unchanged.  We find a similar result if we use OLS for the data concerned with retake exams.

\section{\label{sec:OtherGaps}Not all Achievement Gaps Changed}
In this appendix we show the calculations leading to our two conclusions that i) the gender gap seemed unaffected by the concepts-first structure and ii) the URM gap seemed unchanged by offereing retake exams.

First, we show that the class organization seems unrelated to the gender gap.  We do this by adding a categorical variable for the students' self-identified (binary) gender to Equation \ref{eqn:HLMModel1}.  $Female$ = 1 if the student identifies as female and = 0 if they identify as male.  We also add in the appropriate interaction term to determine if any gender gap is different for the concepts-first class.  In other words, we fit the normalized final exam scores ($NFnlExam$) with the following model:
\begin{multline}
NFnlExam = b_0 + b_{CncptFrst}CncptFrst \\ + b_{URM}URM \\+ b_{URM*CncptFrst}(URM*CncptFrst) \\ + b_{Female}Female \\+ b_{Female*CncptFrst}(Female*CncptFrst)
\label{eqn:HLMModel2a}
\end{multline}

The results of our HLM fit to equation \ref{eqn:HLMModel2a} are shown in Table \ref{tab:tab8}.  One sees that there is a gender gap ($b_{Female}$) of about 0.33 standard deviations and that the gap is not significantly different for the concepts-first structured class, $b_{Female*CncptFrst}$ has $P = 0.8$.  The other coefficients are essentially unchanged from their values in Table \ref{tab:tab2}.

\begin{table}[htbp]
\caption{The coefficients from an HLM fit to equation \ref{eqn:HLMModel2a} are shown along with their standard errors, z-statistics, and P-values.  Included are $N=633$ students in 4 classes.}
\label{tab:tab8}
\begin{tabular}{c c c c c}
\textbf{Coeff.} & \textbf{Value} &\textbf{Error} & \textbf{z-statistic}
& \textbf{P-value}\\ 
\hline
$b_{ConcptFrst}$ & -0.18 & 0.21 & -0.88 & 0.376 \\
$b_{URM}$ & -0.82 & 0.12 & -6.72 & $<10^{-3}$ \\
$b_{URM*ConcptFrst}$ & 0.83 & 0.26 & 3.23 & 0.001 \\
$b_{Female}$ & -0.33 & 0.10 & -3.31 & 0.001 \\
$b_{Female*ConcptFrst}$ & -0.04 & 0.20 & -0.22 & 0.825 \\
$b_{0}$ & 0.24 & 0.10 & 2.26 & 0.024 \\
\end{tabular}
\end{table}

Finally, we show that the retake exam organization seems unrelated to the URM racial/ethnic gap.  We do this by adding a categorical variable for the students' self-identified ethnicity to equation \ref{eqn:HLMModel4}.  As before, $URM$ = 1 if the student identifies as a member of a racial/ethnic group recognized by the APS as being underrepresented in physics and = 0 if they don't.  We also add in the appropriate interaction term to determine if any URM gap is different for the retake classes.  In other words, we fit the students' grades ($CourseGrade$) with the following model:
\begin{multline}
CourseGrade = b_0 + b_{Retake}Retake \\ + b_{Female}Female \\+  b_{Female*Retake}(Female*Retake) \\ + b_{URM}URM \\+  b_{URM*Retake}(URM*Retake)
\label{eqn:HLMModel4a}
\end{multline}

The results of our HLM fit to equation \ref{eqn:HLMModel4a} are shown in Table \ref{tab:tab9}.  One sees that there is a URM gap ($b_{URM}$) of about 0.34 grade points and that the gap is not significantly different for the retake exam classes, $b_{URM*Retake}$ has $P = 0.523$.  The other coefficients are essentially unchanged from their values in Table \ref{tab:tab5}.

\begin{table}[htbp]
\caption{The coefficients from an HLM fit to equation \ref{eqn:HLMModel4a} are shown along with their standard errors, z-statistics, and P-values.  Included are $N=12,649$ students in 52 non-retake classes and $N = 606$ students in 4 retake classes.}
\label{tab:tab9}
\begin{tabular}{c c c c c}
\textbf{Coeff.} & \textbf{Value} &\textbf{Error} & \textbf{z-statistic}
& \textbf{P-value}\\ 
\hline
$b_{Retake}$ & 0.31 & 0.14 & 2.19 & 0.028 \\
$b_{Female}$ & -0.21 & 0.015 & -13.50 & $<10^{-3}$ \\
$b_{Female*Retake}$ & 0.230 & 0.072 & 3.20 & 0.001 \\
$b_{URM}$ & -0.34 & 0.019 & -17.82 & $<10^{-3}$ \\
$b_{URM*Retake}$ & 0.052 & 0.082 & 0.64 & 0.523 \\
$b_{0}$ & 3.130 & 0.037 & 85.24 & $<10^{-3}$ \\
\end{tabular}
\end{table}

\section{\label{sec:TwoInstr}Retake Classes for Each of the Two Instructors}
Our argument in this paper is that changing the structure of a course may remove equity gaps without changing the course's topics covered, level of the presentation of the material, or the types of exams that the students take.  One of our arguments compared courses that did not offer retake exams with courses that did offer them and showed that, on average, the gender gap disappeared for the courses offering retake exams.  Though these results are evidence in support of our conclusion without further discussion, one might still ask whether the course changes that were important here were the instructors rather than the exam retakes.  In this appendix we calculate i) the gender gaps in courses taught by each of these two instructors over the five years preceding the retake classes to show that each instructor's non-retake courses had significantly non-zero gender gaps and ii) gender gaps for the retake classes to show that each instructor's retake classes were consistent with zero gap.  Because this held for both instructors, we could view the final course as a replication of the original three trial courses.

The instructor who taught the first three retake courses also taught eight similar courses in the previous five years.  The gender gaps in these courses, determined using HLM with equation \ref{eqn:HLMModel3}, are shown in Table \ref{tab:tab14}.  This instructor's courses with retakes had an average gender gap consistent with no gap, P = 0.567.  Their courses without retakes had an average gender gap inconsistent with zero gap, P = 0.001.  The gender gap in this instructor's non-retake courses seems to have been smaller than the course average of -0.21 (see Fig. \ref{Fig2}) even without retakes but the gender gap in their retake courses was still much closer to, and consistent with, zero.  The second instructor taught the fourth retake course and also taught two similar courses in the previous five years.  Their courses with retakes had an average gender gap consistent with zero gap, P = 0.303.  Their courses without retakes had an average gender gap inconsistent with zero gap, P = 0.002.  The gender gap in this second instructor's non-retake courses was about the same size as the overall course average without retakes but was still consistent with zero in their retake classes.

\begin{table}[htbp]
\caption{Gender gaps, as determined by HLM analysis (Eq. \ref{eqn:HLMModel3}.  The errors are standard errors.}
\label{tab:tab14}
\begin{tabular}{c c c c c c}
\textbf{Instructor} & \textbf{Group} & \textbf{N} & \textbf{GenGap} &\textbf{Error} & \textbf{P-value}\\
\hline
First & Regular & 1458 & -0.124 & 0.036 & 0.001 \\
Instructor & Retake & 401 & -0.04 & 0.06 & 0.567 \\
\hline
Second & Regular & 686 & -0.189 & 0.062 & 0.002 \\
Instructor & Retake & 209 & 0.12 & 0.11 & 0.303 \\
\end{tabular}
\end{table}

\section{\label{sec:Asian}Analysis of students with Asian ethnicities}
In our original paper on the concepts-first course \cite{Webb2017} we compared the concepts-first course, Lecture I, with the regular course, Lecture II, taught by the same lecturer using the same course materials. For each demographic group large enough to allow a statistical analysis in this particular comparison the students in the concepts-first class performed either better than or statistically the same as their peers from the same group in the regular class. Two of these larger-group comparisons were for students with Chinese ethnicity and students with ethnicity from the Indian subcontinent. We have noted in a previous paper \cite{Paul2022} that aggregating students with Asian ethnicities risks losing important information about the disaggregated groups and that students with different Asian ethnicities seem to experience very different grade gaps \cite{Paul2022}. Nevertheless, in Ref. \cite{Shafer2021} Shafer et. al. show that the aggregated demographic group consisting of students with Asian ethnicities experience a negative grade gap in an introductory physics course similar to our course so, to compare specifically with that paper, we will include that demographic group in our modeling of concepts-first instruction.

We have already included URM status and gender in our model and both of these groups have non-negligible grade gaps so we will continue to include them.  We use Asian = 1 for students with East Asian, Southeast Asian, and South Asian ethnicities and Asian = 0 for the rest of the students.  The 633 students in our dataset can be divided into exactly 3 groups: 92 URM students, 324 Asian students, and 217 white students.  This allows us to control for gender and estimate the effects of URM status and Asian status with the result that the constants in Eq. \ref{eqn:HLMModel6} which appear to be non-ethnic and non-racial actually correspond to white male students.  Our HLM model is
\begin{multline}
NFnlExam = b_0 + b_{CncptFrst}CncptFrst + b_{URM}URM \\+ b_{URM*CncptFrst}(URM*CncptFrst) \\ +
b_{Female}Female \\ + b_{Female*CncptFrst}(Female*CncptFrst) \\
+ b_{Asian}Asian \\ + b_{Asian*CncptFrst}(Asian*CncptFrst)
\label{eqn:HLMModel6}
\end{multline}

so the first two coefficients, $b_0$ and $b_{CncptFrst}$ correspond to white male average in the regular courses and change in white male average in the concepts-first course.

The results of our HLM fit to Eq. \ref{eqn:HLMModel6} are shown in Table \ref{tab:tab15}.  One sees that our previous findings still hold and, in addition, the coefficient $b_{Asian}$ shows us that Asian students had slightly lower final exam grades than white students in the regular classes and $b_{Asian*CncptFrst}$ shows us that that deficit was not changed in the concepts-first class.  So we see that same effect noted by Shafer et. al. \cite{Shafer2021}.  We found a similar issue in Ref. \cite{Paul2022} for the series of physics courses for bioscience students.  There we saw \cite{Paul2022} that there were grade-scale dependent racial/ethnic grade differentials after controlling for physics understanding in the course (i.e. we did not use ``preparation'' metrics as controls).  Specifically, white students received a grade advantage (relative to other racial/ethnic groups) under 4-point scale grading and a somewhat larger grade advantage under percent-scale grading.  Students of Asian ethnicities received little or no advantage (relative to other racial/ethnic groups) under either grade scale, and students from underrepresented groups received significant grade penalties under percent-scale grading after controlling for physics understanding.

\begin{table}[htbp]
\caption{The coefficients from an HLM fit to equation \ref{eqn:HLMModel6} are shown along with their standard errors, z-statistics, and P-values.  Included are $N=633$ students in the four classes (one concepts-first class and four regular classes).}
\label{tab:tab15}
\begin{tabular}{c c c c c}
\textbf{Coeff.} & \textbf{Value} &\textbf{Error} & \textbf{z-statistic}
& \textbf{P-value}\\ 
\hline
$b_{Asian}$ & -0.448 & 0.096 & -4.65 & $<10^{-3}$ \\
$b_{Asian*CncptFrst}$ & 0.0002 & 0.187 & 0.00 & 0.999 \\
$b_{URM}$ & -1.09 & 0.13 & -8.22 & $<10^{-3}$ \\
$b_{URM*CncptFrst}$ & 0.88 & 0.27 & 3.24 & 0.001 \\
$b_{Female}$ & -0.292 & 0.097 & -3.02 & 0.003 \\
$b_{Female*CncptFrst}$ & -0.068 & 0.198 & -0.34 & 0.731 \\
$b_{0}$ & 0.51 & 0.13 & 3.78 & $<10^{-3}$ \\
$b_{CncptFrst}$ & -0.23 & 0.26 & -0.89 & 0.376 \\
\end{tabular}
\end{table}
We can again use pre-FCI and calculus grades as control variables in this analysis, with a caution that these variables have an unknown amount of bias built into them.  Controlling for these two variables we find that $b_{Asian}$ is reduced by about a factor of three to -0.162 $\pm$ 0.070 with P = 0.020 and the effect of the concepts-first class is still negligible (P = 0.525).

We can also compare students with Asian ethnicities with their peers for both retake and non-retake classes.  The model we use is 
\begin{multline}
CourseGrade = b_0 + b_{Retake}Retake \\+ b_{Female}Female +  b_{Female*Retake}(Female*Retake)\\ + b_{URM}URM +  b_{URM*Retake}(URM*Retake)\\ +
 b_{Asian}Asian + b_{Asian*Retake}(Asian*Retake)
\label{eqn:HLMModel8}
\end{multline}
The results of our HLM fit to this model is shown in Table \ref{tab:tab17}.  Again we find that students with Asian ethnicities receive slightly lower grades (an average of 0.145 grade points lower) than white students and URM students received still lower grades. The grade penalty seen by students of Asian ethnicities did not significantly change (P = 0.300 for $b_{Asian*Retake}$) in the retake classes.  We can again use GPA as a control variable in this analysis, with our standard caution that this variable has an unknown amount of bias built into it.  Controlling for GPA we find that $b_{Asian}$ is reduced by more than a factor of three to -0.041 $\pm$ 0.014 with P = 0.002 and the effect of the retakes is still negligible (P = 0.210).
\begin{table}[htbp]
\caption{The coefficients from an HLM fit to equation \ref{eqn:HLMModel8} are shown along with their standard errors, z-statistics, and P-values.  Included are $N=606$ students in the four retake classes $N=12,649$ students in 52 non-retake classes.}
\label{tab:tab17}
\begin{tabular}{c c c c c}
\textbf{Coeff.} & \textbf{Value} &\textbf{Error} & \textbf{z-statistic}
& \textbf{P-value}\\ 
\hline
$b_{Asian}$ & -0.145 & 0.017 & -8.62 & $<10^{-3}$ \\
$b_{Asian*Retake}$ & 0.086 & 0.084 & 1.04 & 0.300 \\
$b_{Female}$ & -0.206 & 0.015 & -13.54 & $<10^{-3}$ \\
$b_{Female*Retake}$ & 0.231 & 0.072 & 3.21 & 0.001 \\
$b_{URM}$ & -0.433 & 0.022 & -19.80 & $<10^{-3}$ \\
$b_{URM*Retake}$ & 0.10 & 0.10 & 1.05 & 0.294 \\
$b_{0}$ & 3.224 & 0.039 & 83.74 & $<10^{-3}$ \\
$b_{Retake}$ & 0.26 & 0.15 & 1.68 & 0.093 \\
\end{tabular}
\end{table}

\bibliography{CourseBias.bib}

\end{document}